\documentclass[a4paper,12pt]{article}
\usepackage[paper=a4paper,left=25mm,right=25mm,top=25mm,bottom=25mm]{geometry} 

\pdfoutput=1
 \usepackage{setspace}
\usepackage{authblk}

\usepackage{fullpage}
\usepackage{parskip}
\usepackage{titlesec}
\usepackage[section]{placeins}
\usepackage[dvipsnames]{xcolor}

\usepackage{breakcites}
\usepackage{lineno}
\usepackage{hyphenat}
\usepackage[all]{nowidow}

\PassOptionsToPackage{hyphens}{url}
\usepackage[colorlinks = true]{hyperref} 

\hypersetup{
  linkcolor  = RedViolet,
  citecolor = Blue!90!black,
  urlcolor   = Aquamarine!80!black,
  colorlinks = true
}

\renewenvironment{abstract}
  {{\bfseries\noindent{\abstractname}\par\nobreak}\footnotesize}
  {\bigskip}
\titlespacing{\section}{0pt}{*3}{*1}
\titlespacing{\subsection}{0pt}{*2}{*0.5}
\titlespacing{\subsubsection}{0pt}{*1.5}{0pt}

\usepackage{etoolbox}
\usepackage{graphicx}
\usepackage[space]{grffile}
\usepackage{latexsym}
\usepackage{textcomp}
\usepackage{longtable}
\usepackage{tabulary}
\usepackage{booktabs,array,multirow}
\usepackage{amsfonts,amsmath,amssymb}
\usepackage{subcaption}
\newif\iflatexml\latexmlfalse

\AtBeginDocument{\DeclareGraphicsExtensions{.pdf,.PDF,.eps,.EPS,.png,.PNG,.tif,.TIF,.jpg,.JPG,.jpeg,.JPEG}}

\usepackage[utf8]{inputenc}
\usepackage[english]{babel}
\usepackage{float}

\usepackage[natbib=true, backend=biber, sorting=nyt, style=apa]{biblatex}
\defcitealias{EUReg2014}{Regulation [EU] No 536/2014}
\defcitealias{UCIrepo}{Kelly et al., n.d.}
\usepackage{csquotes}

\usepackage{orcidlink}

\begin{document}
\title{On \lq\lq Confirmatory'' Methodological Research in Statistics and Related Fields}

\def\correspondingauthor{\footnote{Corresponding author, email: \href{mailto:boulesteix@ibe.med.uni-muenchen.de}{boulesteix@ibe.med.uni-muenchen.de}, Institute for Medical Information Processing, Biometry, and Epidemiology, LMU Munich, Marchioninistr.\ 15, 81377 Munich, Germany.}}
\author[1,2]{F.~J.~D. Lange\orcidlink{0009-0000-2000-1527}}
\author[1]{Juliane C. Wilcke\orcidlink{0009-0005-7675-6669}}
\author[3]{Sabine Hoffmann\orcidlink{0000-0001-6197-8801}}
\author[1,2]{Moritz Herrmann\orcidlink{0000-0002-4893-5812}}
\author[1,2]{Anne-Laure Boulesteix\correspondingauthor{}\orcidlink{0000-0002-2729-0947}}
\affil[1]{Institute for Medical Information Processing, Biometry, and Epidemiology, Faculty of Medicine, LMU Munich, Munich, Germany}
\affil[2]{Munich Center for Machine Learning (MCML), Munich, Germany}
\affil[3]{Department of Statistics, LMU Munich, Munich, Germany}
\vspace{-1em}
  \date{}
\maketitle

\vspace{-1em}

\begin{abstract}
Empirical substantive research, such as in the life or social sciences, is commonly categorized into the two modes {\it exploratory} and {\it confirmatory}, both of which are essential to scientific progress. The former is also referred to as hypothesis-generating or data-contingent research, while the latter is also called hypothesis-testing research. In the context of empirical methodological research in statistics, however, the exploratory--confirmatory distinction has received very little attention so far. Our paper aims to fill this gap. First, we revisit the concept of empirical methodological research through the lens of the exploratory--confirmatory distinction. Second, we examine current practice with respect to this distinction through a literature survey including 115 articles from the field of biostatistics. Third, we provide practical recommendations toward a more appropriate design, interpretation, and reporting of empirical methodological research in light of this distinction. In particular, we argue that both modes of research are crucial to methodological progress, but that most published studies---even if sometimes disguised as confirmatory---are essentially exploratory in nature. We emphasize that it may be adequate to consider empirical methodological research as a continuum between ``pure'' exploration and ``strict'' confirmation, recommend transparently reporting the mode of conducted research within the spectrum between exploratory and confirmatory, and stress the importance of study protocols written before conducting the study, especially in confirmatory methodological research.
\end{abstract}

\sloppy
\vspace{-1em}
\enlargethispage{\baselineskip}


\section{Introduction}

The ultimate goal of methodological research should be to develop methods that will eventually be used in practice, for example, because they solve practical problems or are better than existing methods. However, in statistics and related computational fields such as machine learning (ML) or bioinformatics, more and more methods are being developed, while at the same time, these developments are not accompanied by a sufficient number of empirical evaluations and comparative studies \citep{sauerbrei2014strengthening,boulesteix2024editorial,Heinze2024}. This proliferation of methods and the lack of evidence to guide the choice between them make it difficult for applied researchers to select appropriate methods for their application, thwarting the goal of methodological research. It has even been suggested that methodological research in statistics and related fields faces a replication crisis \citep{Boulesteix2020, Cockburn2020, Gundersen2023, Hutson2018, lohmann2022ten, Niessl2022, Niessl2024, Pawel2024, Pineau2021}.

In this paper, we argue that this problematic situation is partly based on a conflation of exploration and confirmation in \emph{empirical methodological research}, which we define as the study of the properties and the behavior of methods under investigation 
by means of simulated or real data. 
Exploratory research does not necessarily involve specified, concrete hypotheses and seeks to identify patterns in observed data, thereby generating hypotheses for future studies. Confirmatory research, on the other hand, necessitates the a priori formulation of well-grounded, testable hypotheses and intends to ascertain the veracity of prespecified hypotheses through rigorous evaluation, often using newly obtained data. In this paper, we argue that recognizing the conceptual and practical differences between these two approaches in the context of methodological research will lead to more efficient method development as well as better guidance for applied researchers on which method to use in their specific applications. 

The missing awareness of the distinction between exploratory and confirmatory research
has been identified as one of the driving factors of the replication crisis in empirical substantive research, where it has been repeatedly demonstrated that many research findings cannot be replicated in 
fields such as psychology \citep{OSC2015}, genetic epidemiology \citep{Border2019}, and preclinical cancer biology \citep{Begley2012, Errington2021}.
In a similar vein, we argue that most empirical methodological research is exploratory, but is presented or mistaken as confirmatory research. Methodological researchers tend to present their results as more reliable than they actually are. More specifically, we argue that it is currently difficult to even assess which studies, analyses, and findings reported in the methodological statistical literature are exploratory or confirmatory in nature. 
Therefore, we want to increase awareness of the distinction between exploratory and confirmatory research, which is just as essential for progress in this field as in others. We advocate for more transparently documenting methodological research using study protocols and for reporting the nature of presented research. Note that, in practice, a single study may have both exploratory and confirmatory objectives, analyses, and results.

In the context of empirical methodological research, as specified above, the distinction between exploratory and confirmatory research has received little attention in the scientific literature so far. It was briefly discussed in a statistical context with respect to simulation studies by \citet{Pawel2024} and \citet{Siepe2024simulation}, 
and strictly confirmatory real-data studies were suggested by \citet{Niessl2022}. The concept of truly confirmatory research has, to the best of our knowledge, not been characterized, examined, or discussed at all in the context of empirical methodological research. 

This paper aims at filling this gap and is structured as follows: After giving some background on the exploratory--confirmatory distinction in science in general and in selected substantive fields (\hyperref[sec:background]{Section~\ref*{sec:background}}), we examine empirical methodological research through the lens of this distinction (\hyperref[sec:empiricalmethres]{Section~\ref*{sec:empiricalmethres}}). We then assess current practices in biostatistical methodological research with respect to this distinction using a literature survey of recent biostatistical articles published in the \textit{Biometrical Journal} and \textit{Statistics in Medicine} (\hyperref[sec:assessment]{Section~\ref*{sec:assessment}}). 
With the aim of improving the reliability of empirical methodological research by making the distinction between exploratory and confirmatory research, we formulate tentative recommendations for authors and other stakeholders in \hyperref[sec:recommendations]{Section~\ref*{sec:recommendations}}. 
In the discussion (\hyperref[sec:discussion]{Section~\ref*{sec:discussion}}), we consider a practical limitation of the exploratory--confirmatory distinction as well as some epistemic difficulties with the concept of confirmatory methodological research. The paper ends with a short conclusion (\hyperref[sec:conclusion]{Section~\ref*{sec:conclusion}}).


\section{The Scientific Method and Exploratory and Confirmatory Research in Substantive Research}
\label{sec:background}

After short preliminary remarks clarifying terminology, we briefly review how the research modes {\it confirmatory research} and {\it exploratory research} have been described in social and life sciences. Then, we address the pivotal role of protocols and preregistration.

\subsection{Preliminary Remarks}

Before we review the exploratory--confirmatory distinction in substantive research, we want to make two clarifying remarks on the terminology used in the remainder of this paper.
First, perhaps stating the obvious, the hypotheses that are either generated or tested in the two modes of research must be \textit{scientific} hypotheses, that is, hypotheses about phenomena one intends to evaluate using a scientific process after stating them \citep{Thompson2023scope}. The evaluation of these hypotheses can be rooted in different scientific approaches 
and is not restricted to particular methods. For example, the use of the word ``hypothesis-testing'' to describe research does not prescribe that a statistical test must be employed, which leads to our following second remark.

During the discussion of the exploratory--confirmatory distinction and the research process in general, terms such as ``hypothesis'', ``testing'', ``prediction'', ``variables'', or ``theory'' naturally come up. For some readers, these may already be associated with certain concepts due to their specific use in statistics (e.g., in the context of null hypothesis testing or prediction modeling). However, in the context of this paper, they are almost exclusively used with a nontechnical, field-independent meaning, referring to overarching scientific concepts and ideas, such that they could be used to describe studies in many disciplines. Thus, unless specifically stated otherwise, these and similar scientific terms are always meant in a broader, nonstatistical sense in the remainder of the paper.
Finally, let us note that we intentionally use the terms exploratory and confirmatory {\it research} and not {\it studies}. That is because a single study may have both exploratory and confirmatory objectives, analyses, and results. 

\subsection{Exploratory and Confirmatory Research in Substantive Disciplines Applying Statistical Methods}
\label{subsec:distinction_elsewhere}

Empirical substantive research is commonly categorized into the two modes exploratory and confirmatory research.
This pair of labels is also commonly referred to as hypothesis-generating versus hypothesis-testing, data-contingent research versus data-independent, or postdiction versus prediction \citep{Groot2014, Munafo2017, Nosek2018}. Collectively, these four pairs of labels broadly convey the difference between the two modes of research. More detailed characterizations of exploratory and confirmatory research have been offered in publications on psychology \citep{Hoefler2022, Wagenmakers2012}, biology \citep{Jaeger1998, Nilsen2020}, preclinical \citep{Dirnagl2016, Dirnagl2020a,Kimmelman2014} and clinical research \citep{ICH1998E9}, educational research \citep{Foster2024}, and linguistics \citep{Roettger2021}, among others. Although addressing audiences in different fields, most of the published definitions share a few key aspects, which we will summarize below. This indicates that there is a sort of consensus with respect to the exploratory--confirmatory distinction in the context of empirical substantive research. It should also be noted that, regardless of the field or assigned labels, the two approaches have distinct but equally valuable purposes, complement each other, and are both essential to scientific progress \citep{NAS2019reproRepli, Nilsen2020, Nosek2018, Roettger2021, Schwab2020, Wagenmakers2012}.

The first key aspect of the exploratory--confirmatory distinction is the hypothesis specification. Confirmatory research necessitates the a priori formulation of well-grounded, testable hypotheses to be evaluated. As \citet{Jaeger1998} point out, these hypotheses ``usually do not spring from an intellectual void but instead are gained through exploratory research'' (p.~S64). Exploratory research itself does not necessarily involve specified, concrete hypotheses, although this does not mean that exploratory research is always a completely hypothesis-free endeavor either. The involved hypotheses may just not be precisely defined or only develop while conducting the exploratory investigations \citep{Kimmelman2014, Schwab2020}.

The second key aspect distinguishing the two modes of research relates to the data that is used. While there are no restrictions on the data that may be used for exploratory research, confirmatory research often requires the collection of new data \citep{Jaeger1998, Nilsen2020,Schwab2020}. At least, the used data should be different from the data that was used to generate the prespecified hypotheses \citep{Hoefler2022}. 
Additionally, in confirmatory research, the hypotheses that are prespecified directly inform the research design, data collection or choice, and planning of the data analysis, which are similarly specified before the start of the study to ensure a valid and rigorous evaluation. Exploratory research, on the other hand, which intends to explore data more freely to identify patterns, cannot and should not be planned in its entirety at the beginning of a project. Therefore, fewer details about the research design will be specified at that time.
However, this does not mean that exploratory research, lacking clear hypotheses, is done aimlessly without any direction or planning. For example, \citet{Stebbins2001} defines exploratory research as a ``broad-ranging, purposive, systematic, prearranged undertaking designed to maximize the discovery of generalizations leading to description and understanding'' (p.~3).

Another way the distinction has been framed is with the ``researcher degrees of freedom'' (RDFs; \citealp{Simmons2011}), which refer to the available flexibility in a study's design, data collection, and analysis. As previously stated, in confirmatory research this flexibility is intentionally restricted (i.e., the RDFs are 
reduced) to provide sound empirical evidence about clear hypotheses. In contrast, in exploratory research, the RDFs are intentionally utilized to maximize the potential for unexpected discoveries \citep{Thompson2020point, Hoefler2022}.

\subsection{HARKing and Preregistration}
\label{subsec:HARKing}

As mentioned in the last subsection, exploratory research usually gives researchers large freedom in their investigation of data as they look for results that could turn into hypotheses. While this is to some degree required for exploration to fulfill its intended purpose, it also invites the use of questionable research practices (QRPs). A QRP that is particularly related to the exploratory--confirmatory distinction is post hoc theorizing or ``hypothesizing after results are known'' (HARKing; \citealp{Kerr1998}). A classical example of HARKing would be a researcher retroactively fitting a theory or hypothesis to an interesting exploratory research finding and reporting the combination as if they had stated the theory or hypothesis before looking at the data (i.e., as a confirmatory finding). We want to emphasize that HARKing may also happen unintentionally due to cognitive biases (e.g., confirmation bias or hindsight bias). However, whether intentional or not, presenting an exploratory finding as confirmatory leads to overconfidence in the result. The risk that this result is a false research finding and, therefore, cannot be replicated is higher than if the research were truly confirmatory \citep{Button2013, Nosek2018}.

To distinguish exploratory and confirmatory research clearly and transparently and to ensure the purely confirmatory nature of a particular piece or part of research, the public, time-stamped registration of study plans prior to collecting or accessing data has been suggested \citep{Wagenmakers2012, Nosek2018}. This practice is called preregistration and is similar to the registration of clinical trials, which has become standard over the past two decades \citep{Angelis2004, Munafo2017}. Preregistration has seen growing adoption across various fields in recent years, particularly in psychology and other social sciences \citep{Simmons2021}. The registration is realized by archiving a document on a public independent registry such as the Open Science Framework, which can be used to register research from all disciplines. The contents and level of detail in preregistration documents can vary, ranging from just a basic study design to comprehensive research protocols \citep{Munafo2017}. By publicly registering hypotheses, study design, methods, and analysis plans before the beginning of a study, preregistration also addresses HARKing. Whether researchers engage in HARKing or other QRPs intentionally or not, preregistration allows for an accessible assessment of their extent by others. They can compare the published analyses and results to the publicly archived study plan, provided the preregistered study protocol is sufficiently detailed.


\section{Empirical Methodological Research Through the Lens of the Exploratory--Confirmatory Distinction}
\label{sec:empiricalmethres}

After defining empirical methodological research precisely (\hyperref[subsec:defempmethres]{Section~\ref*{subsec:defempmethres}}), we outline the concept of confirmatory and exploratory research in this context (\hyperref[subsec:definition]{Section~\ref*{subsec:definition}}), particularly addressing the case of articles presenting new methods (\hyperref[subsec:new]{Section~\ref*{subsec:new}}).

\subsection{Definition of Empirical Methodological Research}
\label{subsec:defempmethres}
As mentioned in the introduction, we define empirical methodological research for the purpose of this paper as the study of the properties and the behavior of computational methods in a broad sense, including statistical methods, using simulated or real data. 
The use of data distinguishes {\it empirical} methodological research from theoretical methodological research, where properties of the investigated methods are derived based on theoretical (mathematical) considerations.

Empirical {\it methodological} research, where data and methods are employed with the aim of advancing the knowledge about the methods, must also be distinguished from research that uses data and methods with the aim of gaining knowledge about the subject matter of the data (e.g., a disease of interest) rather than about the methods. 
Note that we use the term \lq\lq methods'' to denote not only analysis approaches, such as statistical tests, regression modeling approaches, or statistical learning algorithms, but also any other procedures contributing to the identification and evaluation of patterns in the data, such as any measures/indices, visualization techniques, or sample size calculation methods.

Furthermore, the concept of empirical methodological research we consider here implies that the investigated methods have already been created and precisely defined, for example, by specifying a complex model and implementing an algorithm to fit it. The term \lq\lq method development'' commonly used in statistics and related fields usually refers implicitly to two processes. The first one is the invention of a (new) method (or method variant), and the second one is the accumulation of knowledge about a method over time once it has been created.
The former process, to which we refer as initial ``method creation'', which may include experimental investigation, takes place before a systematic empirical evaluation. 
Consequently, this kind of method development does not fall under the umbrella of empirical research discussed and assessed in this paper, even though informal partial evaluations conducted (but not reported) by the method's creator often guide this process. 
The latter process, termed ``long-term method development'' for our purposes, however, {\it must} involve empirical evaluations to gain knowledge about a method and is ideally realized across multiple publications, including some without the involvement of the method's creator(s). 
The creation of a method and its first empirical evaluation (e.g., based on a few benchmark datasets) are typically reported in a single paper, classified as a phase~I or phase~II paper according to Heinze et al. (\citeyear{Heinze2024}; see their paper for an in-depth discussion of the notion of \lq\lq phases of methodological research''). However, even when reported in the same work, only the empirical evaluation of the created method falls within the scope of the present paper; the creation of the method does not.

\subsection{Exploratory and Confirmatory Methodological Research}
\label{subsec:definition}

We suggest adopting the definitions of exploratory and confirmatory research outlined in \hyperref[sec:background]{Section~\ref*{sec:background}} for the context of empirical methodological research.
Exploratory research does not necessarily involve a specified hypothesis. It can take numerous different forms and includes first evaluations in articles presenting new methods (see \hyperref[subsec:new]{Section~\ref*{subsec:new}} for more details) or comparison studies performed without any specific hypothesis in mind. 
Confirmatory studies are specifically designed to evaluate prespecified hypotheses about {\it existing} methods and their properties---including but not limited to their performance. Replication studies are an important special case of confirmatory research: the hypotheses and settings considered in a replication study are the same as those considered in the previous study that the authors aim to replicate. However, hypotheses considered in confirmatory research may alternatively be based on prior knowledge from other sources, theoretical results, preliminary investigations, or exploratory research with a completely different design---in which case the confirmatory study cannot be considered a replication study. By virtue of detailed planning as well as careful and focused study design, these well-powered studies intend to provide sound empirical evidence and allow researchers to validate or refute specific claims.  


\begin{table}[h]
    \centering
    \begin{tabular}{llcc} 
    \hline
        && Exploratory & Confirmatory\\ 
        \hline
        \multicolumn{4}{l}{\it Research process}\\
        & Aim of developing and adapting methods & +++ & +\\
        & Focus on precise hypotheses and strong evidence & + & +++\\
        & Detailed research protocol before study onset & + & +++\\
        & Sample size considerations for real-data studies & + & +++\\
        & Statistical tests or justifications & + & +++\\
        \multicolumn{4}{l}{\it Reporting}\\
        & Preregistration & + & +++\\
        & Transparent and detailed reporting & +++ & +++\\
        & Computational reproducibility & +++ & +++\\
        \multicolumn{4}{l}{\it Conceptual features}\\
        & Neutrality & + & +++\\
        & Internal validity (control of biases) & ++ & +++\\
        & External validity (generalizability) & -- & ++\\
        & Sensitivity (finding what might work) & +++ & +\\
        & Specificity (weeding out false positives) & + & +++\\ 
        \hline
    \end{tabular}
    \caption{Comparison of exploratory and confirmatory empirical methodological research. Adapted from Table~1 by \citet[p.~77]{Dirnagl2020a} about preclinical study designs. The \lq\lq +'' and \lq\lq --'' signs indicate the extent to which the corresponding features should be emphasized or sought after in exploratory and confirmatory research.
    }
    \label{tab:table_comparison}
\end{table}


As should be clear by now, confirmatory research does not mean testing either hypotheses or previous empirical findings simply by repeating exploratory research \citep{Dirnagl2020a} using, say, a different seed or slightly different settings (for simulations) or a different collection of datasets (for real-data-based research). Instead, the two modes of research require different study designs along with different research activities, precisely because they serve different purposes: discovery versus confirmatory testing. 
A tentative overview of the differences and similarities between exploratory and confirmatory empirical \enlargethispage{\baselineskip}
methodological research is given in \hyperref[tab:table_comparison]{Table~\ref*{tab:table_comparison}}. While most of the features discussed in this table are highly desirable for one or both modes of research, it will be seen in \hyperref[sec:assessment]{Section~\ref*{sec:assessment}} that, for various reasons, they have been given only limited attention in empirical methodological research to date.

Hypotheses in empirical methodological research may be of different types. They may concern a single method or, most commonly, several methods to be compared in some way. Examples of potential types of hypotheses in confirmatory research can be found in \hyperref[tab:hypotheses_examples]{Table~\ref*{tab:hypotheses_examples}}. 
For all these hypotheses, regardless of whether comparative or not, the researcher also has to specify the considered setting in some way. For example, which type of real datasets, which range of settings for simulation studies, or which benchmark database is considered? Furthermore, all concepts involved in the hypotheses (e.g., \lq\lq similar'', \lq\lq performs better'') have to be precisely defined depending on the context.


\begin{table}[h]
\centering
\begin{tabulary}{0.95\textwidth}{L>{\hangindent=0.84em}L}
\hline
\multicolumn{2}{l}{\it Hypotheses on a single method}  \\ 
& Estimator~A is not unbiased if condition~C is not fulfilled.\\
& Regression approach~A needs a sample size $>N$ to converge.\\ 
\multicolumn{2}{l}{\it Comparative hypotheses}    \\ 
& Methods~A and~B yield similar results. \\
& Methods~A and~B have similar performances. \\
& Method~A is more robust than method B in setting~X, and vice versa for setting~Y. \\
& Method~A always performs better than or similarly to method~B.\\
& Method~A is on average X times faster than method~B. \\ \hline
\end{tabulary}
   \caption{Potential generic hypotheses to be considered in confirmatory empirical methodological research.}
   \label{tab:hypotheses_examples}
\end{table}


When embarking on confirmatory research, a detailed research protocol with the precise study design, including the specific hypotheses to be investigated and the experimental plan, as well as the analysis plan must be written before conducting the study. This helps to decrease unwanted data dependency and to increase consistency between all research parts as well as internal validity so that the research actually addresses what is of interest. For research based on real datasets, the protocol should also contain sample size considerations (i.e., a justification of the number of considered real datasets) to ensure adequate statistical power for detecting the effects of interest. In the same vein, the number of simulation repetitions per setting in a simulation study should be chosen to ensure sufficient precision of the results. 

Whereas some form of inference or at least clearly defined objective criteria are essential in confirmatory research, exploratory research often relies much more on descriptive statistics. 
Note that the use of confirmatory data analysis methods, such as hypothesis testing, for analyzing the results of a comparison study does not necessarily make the study confirmatory. The other way around, exploratory data analysis methods may be used within confirmatory research. As can be seen from \hyperref[tab:table_comparison]{Table~\ref*{tab:table_comparison}} and will be further discussed in the next subsection, the classification of research as confirmatory or exploratory is not solely dependent on the nature of the methods used to analyze the results. It is also important to note that statistical inference in the context of methodological research is a complex exercise, which can take different forms not limited to simple hypothesis tests. Depending on the study design, various approaches can be used, such as simple parametric or non-parametric tests \citep{benavoli2014bayesian,benavoli2017time,Boulesteix2015statistical,demsar2006statistical,eisinga2017exact}, regression models, including linear mixed models \citep{eugster2012domain} and factorial ANOVA \citep{chipman2022lets,van2023white}, and the Bradley--Terry model commonly used in psychology \citep{eugster2014psycho}.  
In any case, precisely defining the population of datasets of interest and the considered hypotheses---along with the associated parameters---is a difficult task that requires careful attention (see the discussion in \citealp{Boulesteix2015statistical}).

All research must be reported transparently and with sufficient detail (e.g., the design of the comparison studies, potential biases, the exploratory or confirmatory character) to allow readers to assess the results and attempt their replication. Moreover, computational reproducibility needs to be ensured, for example, by providing open data and code that is understandable and reproducible. While these aspects are important for all modes of research, they are particularly crucial 
for confirmatory research. As outlined above, the credibility of confirmatory research is higher if the research protocol is preregistered.

Furthermore, confirmatory research should be conducted in a neutral manner, while this aspect is less crucial for exploratory research. Neutrality implies that measures are taken---ideally in all phases of the study (design, implementation, interpretation, reporting)---to avoid a biased assessment of methods due to the authors' preferences, knowledge, or skills (see \hyperref[subsec:new]{Section~\ref*{subsec:new}} for a discussion of neutrality focused on articles presenting new methods). Biases should also be avoided, reduced, or controlled as much as possible by measures such as randomization, blinding, and prespecified inclusion and exclusion criteria so that adequate inferences can be drawn.
Unlike findings from exploratory empirical methodological research, those from confirmatory research need to be sufficiently general, meaning that the context in which the confirmatory results are supposed to hold should be defined carefully. Typically, inferences are made for more than the specific investigated dataset or simulation setting, which might involve extensive evaluations of methods using many realistic simulation scenarios, data-generating mechanisms, or datasets. 
Finally, when using statistical tests to analyze the results of their methodological studies, researchers working in exploratory mode aim to keep the risk of false negatives in check (high sensitivity), not wanting to miss potential solutions, whereas researchers working in confirmatory mode insist on a low risk of false positives (high specificity) because they should not draw further resources or be applied in practice.

\subsection{The Case of Articles Presenting New Methods}
\label{subsec:new}

Given that much of empirical methodological research appears to be published in articles presenting a new method (see the results of our survey in 
\hyperref[subsec:survey]{Section~\ref*{subsec:survey}}), that is, in articles including method creation contents, such studies are given particular consideration here.
We make the argument that, in most cases, claims about new methods based on these experiments are the result of exploratory rather than confirmatory research for the following reasons:

First, we argue that, without any previous empirical evidence about the newly proposed method, one probably lacks the information needed to plan a study in a confirmatory manner. 
Second, even if this likely practical limitation would not exist, the fact that the published empirical study is conducted by the creator(s) of a method gives rise to a number of potential issues that weaken the validity of findings and claims about said method. These issues are related to the non-neutrality of the studies due to creators having a preference for or being more familiar with their new method relative to the other methods in their study \citep{Boulesteix2013neutral,Boulesteix2017}. Needless to say, such non-neutrality, which may also have only subconscious influence, can lead to bias in every aspect of a study, from planning and design to analysis and reporting.

Third, another potential problem could be described as data leakage between the method creation process and the published empirical study. It occurs when the datasets (or simulation scenarios) analyzed in the empirical study were already utilized in some form during the method creation, for example, in early experiments that guided the creation process, essentially meaning that the same data are used for both method creation and method evaluation \citep{jelizarow2010over}. If a dataset used to develop a method is also used to evaluate its performance, the resulting evaluation is likely to be overly optimistic. This is because the method becomes overly tailored to the specific characteristics of that dataset and may not generalize well to others. 

Data leakage can also take more subtle and indirect forms. Without involving the datasets directly, leakage may also be present if a method creator had prior knowledge about the datasets (e.g., their characteristics or results of existing studies), which likely biased their method creation, even if it was not consciously utilized \citep{salzberg1997comparing}. While papers exclusively investigating existing methods (i.e., methods for which an evaluation has been published) can certainly also have neutrality issues, claims therein are potentially less exploratory as information about the methods was already available, and authors often have a weaker personal connection to the methods.

In summary, the above-mentioned problems impact the validity of evaluations presented as part of papers introducing new methods to such a degree that they cannot fulfill the criteria characterizing confirmatory research. Moreover, even if one is aware of their existence, these problems cannot always be averted from the creator’s perspective, let alone be detected by readers. Therefore, we maintain that an evaluation of a method newly proposed in the same article should be considered provisory and, with few exceptions, cannot be considered as confirmatory research. 
That is not to say that an article presenting a new method could never contain any confirmatory claims. It could, for example, do so as the result of secondary research aims about existing methods. 
However, given that articles about new methods are naturally very focused on the proposed method, we would argue that such findings are more the exception than the rule.


\section{Current Practices in Biostatistical Methodological Research}
\label{sec:assessment}

In this section, we assess current practices in biostatistical methodological research with respect to the exploratory--confirmatory distinction through a literature survey (\hyperref[subsec:survey]{Section~\ref*{subsec:survey}}) and discuss the high prevalence of exploratory research (\hyperref[subsec:empirical]{Section~\ref*{subsec:empirical}}) as well as the notion of pseudo-confirmatory research (\hyperref[subsec:pseudo]{Section~\ref*{subsec:pseudo}}).

\subsection{A Survey of Recent \textit{Biometrical Journal} and \textit{Statistics in Medicine} Articles}
\label{subsec:survey}
To get an up-to-date picture of common practices in methodological research in biostatistics, particularly with respect to the concepts discussed in this article, we conducted a short survey of articles published in the biostatistical journals \textit{Biometrical Journal} (\textit{BiomJ}) and \textit{Statistics in Medicine} (\textit{StatMed}) in 2023. The survey was conducted by the first author, who consulted the last author in cases of uncertainty. It focused on the extraction of features that can, to a large extent, be considered as objective, such as whether a new method (or method variant) is introduced, whether words such as \lq\lq exploratory'' or \lq\lq confirmatory'' are used, and whether the authors refer to research hypotheses. In particular, the objective of the survey was not to definitively classify the articles as either confirmatory or exploratory, as such classification would inevitably involve subjectivity---especially since our paper is the first to explicitly introduce this distinction, and authors could not have been expected to refer to it explicitly. 

Starting with the most recent issues published at the time of the survey (issue 7 from volume 65 of \textit{BiomJ}, issue 22 from volume 42 of \textit{StatMed}) and excluding special issues, we screened issues in reverse chronological order until we reached at least 50 articles containing empirical methodological research from each journal. 

We extracted all eligible articles from the issue with which this target was met, resulting in slightly more than 50 articles per journal in our survey. The data on which the results presented in this section are based (i.e., the spreadsheet documenting the survey) is provided in the Supporting Information.
Ignoring editorials, comments, tutorials, book reviews, corrections, and letters, we extracted 122 research articles (61 from each journal), of which 115 contain empirical methodological research as defined above ($57$ from \textit{BiomJ} and $58$ from \textit{StatMed}). In the vast majority of the surveyed articles ($100/115=87\%$), only simulated data was used to study the investigated method(s). Eleven articles presented real-data methodological studies, and $4$ articles used both simulated data and real data. In articles where the methodological study did not involve real data, numerical results were almost always augmented with an illustration of the method(s) on one or two real datasets. 

In no article that we surveyed do the authors use the term ``exploratory'' or ``confirmatory'' in reference to any part of the presented research. Although we found several examples where the verb ``explore'' was used to describe authors' intentions or where ``confirm'' was used during the interpretation of results, the research itself was never explicitly stated to be exploratory or confirmatory. This is in line with our general impression that the exploratory--confirmatory distinction is currently not being made or considered by researchers in the field of methodological research. 
Only one article explicitly states hypotheses or predictions. Specifically, \citet{Chipman2023} ``hypothesized that (H1) the optimal, SMR [sequential matched randomization] fixed matching threshold would be sensitive to covariate distribution, covariate association with outcome, and sample size and that (H2) each extension, individually and potentially collectively, would improve covariate balance and estimator efficiency'' (p.~3985) and explicitly refer to each hypothesis when they describe the results. However, it should be noted that the authors a) do propose some of the evaluated approaches in the same article, which makes non-neutrality more likely, b) themselves describe the investigated simulation scenarios as ``simplified settings'' throughout the article, do not justify them and also do not comment on how realistic they are, and c) do not specify on what their hypotheses were based. Therefore, even though there were explicit hypotheses, the simulation study in question, in our view, does not constitute confirmatory research. 

We notice that authors frequently report their results to be ``as expected'', and we also find a few examples of researchers describing their expectations before they report their empirical study. Both of these indicate that methodological researchers do have hypotheses or predictions before analyzing their simulated or real data, but generally do not report them explicitly.

Unsurprisingly, it is difficult to judge whether a given paper contains exploratory or confirmatory research (or both). This is not only due to the absence of explicit descriptions in that regard or explicit hypotheses. It can also be challenging because method development, newly proposed methods, or articles presenting new methods, which represent a large part of the methodological literature, have not been discussed in relation to the two modes of research. Therefore, we elaborate on this particular topic in the following subsection.

Finally, let us note that in 92\% (106) of the articles in the survey, the evaluated methods, procedures, or method variants include one that was proposed in the same article. Conversely, just 9 articles (6 from \textit{BiomJ} and 3 from \textit{StatMed}) in our survey covered exclusively existing methods, and all of them compared multiple methods. This may not be a surprising finding since both journals emphasize new methodology in the description of their scope. However, in similar surveys of the methodological literature, which considered only real-data studies, the proportion of articles that both present and evaluate a new method was only slightly lower, e.g., 83\% \citep{norel2011self} or 78\% \citep{Boulesteix2013neutral}.

\subsection{Empirical Methodological Literature Dominated by Exploratory Research?}
\label{subsec:empirical}

Given the rarity of explicit hypotheses and the predominance of articles focused on the evaluation of new methods, one might conclude that published empirical methodological research is almost always exploratory. 
Besides the already mentioned observations, there is further evidence indicating that this is the case. 
To begin with, there is the fact that the results of statistical simulation studies are usually analyzed only descriptively using exploratory data analysis methods (e.g., graphs and tables), and inferential analyses of simulation findings are rare. 
In their survey of 677 statistical research articles with simulation studies (published across six statistical journals between 1985 and 2012, including 194 and 187 articles from 2009 and 2012, respectively), \citet{Harwell2018survey} found that in 99.9\% of them, the authors limited the analysis of their study’s results to graphs and tables. 
While---as outlined above---it would certainly be over-simplistic to judge the nature of a study based solely on the employed analysis methods, we argue that confirmatory claims should be at least partly based on some sort of inferential analysis in a broad sense.

As far as real-data studies are concerned, two common practices suggest that they generally do not meet the standards of confirmatory research: a) most studies use only one or a few datasets, and b) due to the limited number of datasets available, most studies at least partially overlap in the datasets they use.
\citet{Macia2013} surveyed 215 ML papers published between 2008 and 2010 and found that over 80\% of studies used either one dataset (25.5\%) or 2--10 datasets (55.5\%). They also reported that in 63.9\% of studies, the datasets were sourced from the popular UCI repository (\url{https://archive.ics.uci.edu/}), with some datasets being studied particularly often (e.g., 15 datasets were used in at least ten of the surveyed papers, five of which even in at least 20 papers). Similarly, in their survey of 55 articles with real-data comparison studies (published between 2010 and 2012), \citet{Boulesteix2013neutral} reported regarding the number of datasets that the median was 5, 
the upper quartile was 7.5, and the maximum was 21. More recently, \citet{liao2021are} surveyed 103 papers published at major ML conferences between 2016 and 2021 and found that more than 65\% of them used three datasets or fewer, while the overall mean number of datasets was 4.1. In the same vein, \citet{koch2021rrr} analyzed the usage of 2063 datasets across 26,691 ML papers published between 2015 and 2020 and found a concentration on a limited number of datasets within different ML communities that has been increasing in the considered period.
Even if one assumes that many studies have prespecified hypotheses, the majority of studies use (very) small numbers of datasets, and the frequent dataset reuse increases the chances of the hypotheses being tested on (partially) the same data that were used to generate them (i.e., data leakage between studies). 

It is difficult, perhaps even impossible, to ascertain whether a finding in the existing literature was the result of confirmatory research, but it is probable that the vast majority of the empirical methodological literature so far has presented nonconfirmatory research. At this point, we want to reemphasize that exploratory research is just as valuable as confirmatory research, and there are also indications that current empirical methodological research is at least not entirely exploratory. 
Even though none of the articles included in our survey have explicitly prespecified hypotheses, many of them contain wording that suggests that the authors did have at least vague hypotheses before conducting their study (e.g., frequent use of \lq\lq as expected'' when describing results). However, considering our previous observations about current practices, it is unlikely that findings in those articles would qualify as confirmatory, even if the authors had explicitly prespecified their expectations as hypotheses. 

Another observation about current methodological research is that large simulation studies comprising various settings are becoming more and more common. Of course, a comprehensive evaluation is only one characteristic of confirmatory research. 
However, at least in terms of extensiveness, existing methodological research may not be as far from confirmatory as one might think. 
Recent editorial initiatives emphasizing the importance of neutral evaluations of existing methods are also suggestive of this increasing awareness. For example, in 2024, the {\it Biometrical Journal} published a special collection entitled \lq\lq Towards Neutral Comparison Studies in Methodological Research'' \citep{boulesteix2024editorial}. 

Ultimately, we cannot say with certainty how prevalent exploratory and confirmatory research are in the methodological literature or if the predominance of exploratory work is as pronounced as we argue. The assessment with respect to the exploratory--confirmatory distinction was---as anticipated---very difficult, in part because the terms exploratory research and confirmatory research are not established for methodological research.
A broader issue was the reporting of methodological studies, where often little to no information or description regarding the epistemic intentions or research mode is given. This is a consequential weakness of most current methodological research, one that not only complicates distinguishing between exploratory and confirmatory works but also impedes 
the readers' ability to calibrate their confidence in the presented results in general. 
Both exploratory and confirmatory research are essential to scientific progress, and it is just as crucial that the reporting transparently conveys what kind of research was conducted. 
Currently, in methodological research, with this information largely missing from the reporting, there is particularly the risk that exploratory results are misperceived as confirmatory findings.

\subsection{Pseudo-Confirmatory Findings}
\label{subsec:pseudo}

We denote as {\it pseudo-confirmatory findings} nonconfirmatory findings that are presented as if they were the result of confirmatory research. Pseudo-confirmatory findings can arise either from research intended as exploratory from the start or from research intended as confirmatory but where QRPs were employed. The term is new, but the phenomenon is known in many fields. Based on our survey, we suspect that pseudo-confirmatory findings are fairly common in empirical methodological research.

Formulating a research hypothesis a posteriori {\it after} conducting exploratory data analyses and pretending it was formulated {\it before} is a form of HARKing (see \hyperref[subsec:HARKing]{Section~\ref*{subsec:HARKing}}). 
Note that, in methodological research, HARKing can take particularly subtle forms, which makes it hard to detect, as outlined by \citet{Bell2021} and in our \hyperref[subsec:new]{Section~\ref*{subsec:new}} describing data leakage between the trial-and-error process characterizing method creation and the first empirical study presented as part of the paper introducing the new method. \citet{hullman2022worst} frame this issue as \lq\lq [c]ommunication concerns [...] includ[ing] tendencies to not report trial and error over the modeling pipeline and evaluation metrics'' (p.~342). 
For research intended as confirmatory in the first place, on the other hand, 
the involved QRPs range from selective reporting to changing the prespecified hypothesis after knowing the result, which can again also be viewed as a variant of HARKing.  

In benchmark studies using real datasets, for example, researchers sometimes engage in post hoc exclusions of certain benchmarked methods or datasets. It is also likely that they sometimes exploit the flexibility of the design of their benchmark and perform many different benchmark variations in the hope of finding the superiority of a particular method \citep{Boulesteix2017, Niessl2022, norel2011self}. Similar mechanisms may be at work in simulation studies \citep{ullmann2023over,Pawel2024}. 
While most authors engaging in QRPs do so unintentionally (i.e., without malicious intention), one may argue that such bad research practices should and could simply be avoided in confirmatory research.

However, discrepancies between the way a study was planned and the way it is reported can also be more subtle and thus harder to avoid. \citet{Siepe2024simulation} state that \lq\lq [s]ome may argue that simulation studies are often conducted at a more exploratory stage of research and therefore do not require as much rigor and transparency [...]. However, many simulation studies are not conducted and reported as exploratory, but rather with the explicit goal of deriving recommendations for the use of methods'' (p.~2). Thereby, they point to the discrepancy between the exploratory mode characterizing many simulation studies and the intended impact of the conclusions that authors draw from these studies. 
We also consider such studies as pseudo-confirmatory, just in a slightly different sense than when considering HARKing and selective reporting.

Our assessment of the biostatistical methodological literature has uncovered blind spots and a fair amount of uncertainty with respect to exploratory versus confirmatory research. To improve this unsatisfactory state, we now turn to a number of measures that researchers and other stakeholders can implement to improve the reliability, reporting, and usefulness of methodological research results---some with little, some with more effort.


\section{Recommendations}
\label{sec:recommendations}

Based on insights from our survey regarding exploratory and confirmatory empirical methodological research, 
we now formulate tentative recommendations for researchers, editors, funders, and other stakeholders, with the aim of improving the reliability of this research in both the short and long run. As should be the case, several of the recommendations go back to features of exploratory versus confirmatory methodological research introduced in \hyperref[subsec:definition]{Section~\ref*{subsec:definition}}. An overview of our recommendations is given in \hyperref[tab:recommendations]{Table~\ref*{tab:recommendations}}, which we discuss in the following two subsections. What may be most striking about this overview is that observing the distinction between exploratory and confirmatory research is relevant to the majority of the recommendations for improving the reliability and reporting of empirical methodological research---either directly or because of differing importance in the two research modes.


\begin{table}
\centering
\begin{tabulary}{0.95\textwidth}{L>{\hangindent=0.84em}L}
\hline
\multicolumn{2}{l}{\it Researchers} \\ 
& Planning and conducting studies: \\
& $\ast$ Acknowledge the exploratory--confirmatory distinction in empirical methodological research \\ 
& $\ast$ Write research protocols, and preregister especially those with confirmatory parts \\
& -- Carefully consider the selection of datasets and data-generating mechanisms, including underlying rationales \\ 
& $\ast$ Undertake efforts to ensure or increase neutrality (especially for confirmatory research)\\ 
& -- Conduct sensitivity analyses \\ 
& Reporting studies: \\
& $\ast$ Specify the nature of (all parts of) the reported research \\ 
& $\ast$ State any hypotheses explicitly (especially for confirmatory research) \\ 
& -- Be transparent about the degree of neutrality and the study process \\ 
& $\ast$ Interpret results according to the nature of the research and report them correspondingly\\ 
& -- Share (comprehensible) computational code and (if possible) data \\
\multicolumn{2}{l}{\it Editors, journals, funders, and other stakeholders} \\ 
& Reporting quality: \\
& $\ast$ Require explicit statements about the nature of research \\ 
& -- In articles with new methods, encourage sections for describing the (usually hidden) creation process \\ 
& -- Establish mandatory statements of neutrality regarding the studied methods \\
& $\ast$ Promote common standards in studies (especially for confirmatory research) \\ 
& $\ast$ Encourage comprehensive research protocols (especially for confirmatory research) \\
& -- Promote standardized reporting and the development of reporting guidelines \\
& -- Require computational code and (if possible) data, and check computational reproducibility \\
& Research directions: \\
& $\ast$ Distinguish between exploratory and confirmatory research, valuing both approaches equally\\
& -- Encourage in particular neutral research about existing methods \\ 
& $\ast$ Promote preregistered confirmatory research projects \\ 
& -- Fund meta-scientific projects examining methodological research itself \\
\hline
\end{tabulary}
\caption{Recommendations for improving the reliability and reporting of empirical methodological research. Starred items are particularly related to the exploratory--confirmatory distinction, either directly or because of their differing importance in the two research modes.}
\label{tab:recommendations}
\end{table}


\subsection{Recommendations for Researchers}

When dealing with the research literature and when performing their own research, it is important that methodological researchers are aware of the distinction between exploratory and confirmatory research, and hence of the benefits and limitations of these two modes of research. More specifically, this means trying to identify the nature of others' research, so that their findings can be interpreted more adequately, as well as considering the distinction when planning and conducting one's own research so that misattributions can be avoided, the respective strengths better exploited, and weaknesses properly addressed. Note, however, that a single piece of research may in practice have both exploratory and confirmatory parts or features, and should therefore be located as a whole somewhere between the exploratory and confirmatory poles of this dimension (see \hyperref[sec:discussion]{Section~\ref*{sec:discussion}} for more on this, including detailed examples). 

Documenting research plans, which are needed to implement any research idea, has various advantages, particularly if the plans are sufficiently detailed. Written research plans are useful for helping to ensure not only a more systematic and consistent approach but also their more faithful realization and reporting, better internal communication, and fewer QRPs. As in substantive disciplines applying statistical methods, research plans in methodological research allow for the assessment of QRPs by others, who can compare the reported results to the study plan, provided it is sufficiently detailed. This, in turn, may dissuade researchers from engaging in QRPs. What is more, written research plans can help to clarify the intended exploratory versus confirmatory nature of the research. In the context of methodological research, the writing of protocols has been advocated and recommended for simulation studies several times over the past 20 years \citep{Burton2006,Morris2019,Smith2010}, and its potential benefits for real-data studies have also been recognized \citep{Boulesteix2017}. Unfortunately, there are no study protocol templates for empirical methodological research as there are for clinical trials and other fields, a notable exception being the recent template by \citet{Siepe2024simulation} for simulation studies in psychology. We therefore suggest a list of essential items for methodological researchers to address in their research protocols (see \hyperref[tab:checklistnew]{Table~\ref*{tab:checklistnew}}). Especially those protocols with confirmatory parts will benefit from preregistration (see \hyperref[subsec:HARKing]{Section~\ref*{subsec:HARKing}}), which may later be supplemented by justified amendments and, at the time of research completion, a list of deviations.


\newcommand{\PreserveBackslash}[1]{\let\temp=\\#1\let\\=\temp}
\newcolumntype{W}[1]{>{\PreserveBackslash\raggedright}p{#1}}
\begin{table}
\centering
\begin{tabular}{W{8.5cm}l}
\hline
\multicolumn{2}{l}{\textit{General information}} \\
\multicolumn{2}{l}{-- Title; authors (with contributions if possible)}  \\
\multicolumn{2}{l}{-- Funding; conflicts of interest}  \\
\multicolumn{2}{l}{-- \begin{tabular}[t]{@{}l@{}}Version number and date of the current protocol version; protocol revision\\ history with version numbers and dates\end{tabular}}  \\
\multicolumn{2}{l}{-- \begin{tabular}[t]{@{}l@{}}Synopsis, incl.\ description of the study's nature
\end{tabular}} \\
\multicolumn{2}{l}{\textit{Introduction}} \\
\multicolumn{2}{l}{-- Brief overview, rationale for the study, and the study’s overall aims} \\
\multicolumn{2}{l}{-- \begin{tabular}[t]{@{}l@{}}Summary of relevant previous work, both published and unpublished \end{tabular}}  \\
\multicolumn{2}{l}{-- Objectives, research questions, and, for confirmatory studies, hypotheses} \\
\textit{Real datasets} & \textit{Simulated datasets}\\
-- Description of the collection of datasets
 & -- Details on the data-generating  \\
-- Inclusion and exclusion criteria & \hspace{0.84em}mechanisms, factors, settings\\
-- Sample size and power considerations & -- Number of simulated datasets \\
-- \begin{tabular}[t]{@{}l@{}}Description of the dataset selection process \\ and its results \end{tabular} & \hspace{0.84em}(repetitions) for each setting \\
\multicolumn{2}{l}{\textit{Study design}} \\
\multicolumn{2}{l}{-- Studied methods and measures for characterizing their behavior} \\
\multicolumn{2}{l}{-- Preprocessing, experimental design, and validation procedure} \\
\multicolumn{2}{l}{-- Parameters, configurations, and hyperparameter tuning details for each method} \\
\multicolumn{2}{l}{\textit{Analysis plan}} \\
\multicolumn{2}{l}{-- Description of planned analyses, separating confirmatory from other analyses} \\
\multicolumn{2}{l}{\quad\quad -- \begin{tabular}[t]{@{}l@{}}For confirmatory analyses (for each hypothesis): Operationalization\\ of hypothesis and definition of evaluation metric; statistical techniques to\\ evaluate hypothesis; inference criteria\end{tabular}}  \\
\multicolumn{2}{l}{-- \begin{tabular}[t]{@{}l@{}}Specification of contingencies and backup analysis plans for common issues in\\ empirical methodological studies (e.g., outliers or non-convergence of algorithms) \end{tabular}}  \\
\multicolumn{2}{l}{-- \begin{tabular}[t]{@{}l@{}}Planned analyses to investigate the results with respect to study design, analysis \\choices, and characteristics of the datasets or data-generating mechanisms \end{tabular}}  \\
\multicolumn{2}{l}{\textit{Software, hardware, and reproducibility}} \\
\multicolumn{2}{l}{-- \begin{tabular}[t]{@{}l@{}}List of software, central packages and dependencies \end{tabular}}  \\
\multicolumn{2}{l}{-- Details on the implementation of the studied methods} \\
\multicolumn{2}{l}{-- \begin{tabular}[t]{@{}l@{}}Details on the hardware and computational power that will be employed \end{tabular}} \\
\multicolumn{2}{l}{\textit{Prior knowledge, neutrality, and study timeline}} \\
\multicolumn{2}{l}{-- \begin{tabular}[t]{@{}l@{}}Known prior work based on the selected datasets or data-generating mechanisms, \\the analyzed measures in that work, and its relation to the planned study \end{tabular}} \\
\multicolumn{2}{l}{-- Prior knowledge about the datasets or data-generating mechanisms themselves} \\
\multicolumn{2}{l}{-- Neutrality statement regarding the investigated methods} \\
\multicolumn{2}{l}{-- Steps taken to enhance or ensure the neutrality of the study (e.g., blinding)}  \\
\multicolumn{2}{l}{-- \begin{tabular}[t]{@{}l@{}}Study timeline (e.g., the extent to which preliminary analyses or simulations have\\ been conducted at the time of writing) \end{tabular}}  \\\hline
\end{tabular}
\caption{Suggested essential items for study protocols for empirical methodological research in statistics and related fields.}
\label{tab:checklistnew}
\end{table}

There are a few further points to which empirical methodological researchers should pay particular attention when planning and conducting studies, depending on the nature of their research. First, the process of selecting datasets and data-generating mechanisms, including the underlying rationales, is often more crucial than it may appear and needs corresponding time and effort, even if this is somewhat tedious. Secondly, neutrality with respect to the investigated method(s) is likely to be beneficial in exploratory research and especially important in research intended as confirmatory, given that bias in the assessment of methods will tend to distort both the results and the reporting of empirical methodological research. Measures to promote neutrality include blinding regarding the methods while they are being evaluated (e.g., implemented in the code or through external analysts) and collaborating with researchers regarding less familiar methods. Thirdly, conducting sensitivity analyses and investigating alternative analysis strategies tend to make results less context-dependent.

Naturally, many of these aspects are similarly relevant when {\it reporting} empirical methodological studies. First of all, specifying the nature of the research greatly facilitates adequate interpretation by other researchers (cf.\ our difficulties in the reported literature survey), including more favorable assessments by reviewers \citep{Jaeger1998}. For mixed research, this should be done for each part, and it may then also be helpful to separate exploratory from confirmatory analyses and findings \citep{Groot2014, Thompson2020point}. Given the apparently high prevalence of pseudo-confirmatory findings in the methodological literature, reflecting on the intended versus realized nature of one's research seems worthwhile for everyone. Paying attention to this nature should then also make it easier for authors themselves to interpret their results according to the nature of the research and to report their findings correspondingly, using appropriate and precise language (e.g., not suggesting a more confirmatory character for findings of analyses intended as exploratory).

We have already emphasized the importance of transparent and detailed reporting elsewhere (see \hyperref[subsec:definition]{Section~\ref*{subsec:definition}}, for example) and do so again here. While it is desirable for all research to explicitly state any hypothesis that was investigated, this is essential for research intended as confirmatory. Transparent reporting of the study process might furthermore involve details on, for example, the aforementioned selection of datasets and data-generating mechanisms, including the rationales behind this selection, or the timeline of data access. In addition, authors may link the degree of their neutrality with respect to the methods investigated or evaluated to, for example, their involvement in the development of the methods or their otherwise acquired prior knowledge of, familiarity with, or preferences for the methods. Finally, sharing both data (if possible) and computational code that is comprehensible and makes the results reproducible is highly recommended to promote credibility and further research.

\subsection{Recommendations for Journals and Other Stakeholders}
\label{subsec:recommendations_journals}

The quality of reporting is more likely to improve if the motivation and responsibility for this change in the research culture lie not only with individual researchers as authors but also with other stakeholders and roles, such as journal editors and reviewers. Regarding the distinction between exploratory and confirmatory research, they could require explicit statements from authors about the nature of the reported research, and they could encourage the already mentioned separation of exploratory and confirmatory parts in publications. Similarly, authors of manuscripts presenting new methods should be encouraged to include sections for transparently describing the method creation process (before any empirical methodological research as defined in \hyperref[subsec:defempmethres]{Section~\ref*{subsec:defempmethres}}), which may include such steps as trial and error or preliminary simulations. Journals could further support the change in reporting culture by providing authors with a reporting checklist regarding the nature of their research as well as other details mentioned in the previous subsection (e.g., hypotheses, neutrality, adequate interpretations), and reviewers with a corresponding assessment checklist. 

Another measure to increase the reporting quality in empirical methodological research is for journals, and possibly funders, to establish mandatory statements of neutrality regarding the studied methods. This would be somewhat akin to positionality statements in the social sciences or conflict of interest statements taken to include personal interests. While assessments of bias may be subjective, they would make authors and readers more aware of this common issue and probably bring to light additional useful information. More generally than this, the promotion of common standards regarding such study features as definitions of applied terms, measurements, and indicators, and dealing with deviations from plans, would ease adequate interpretation of findings, synthesis of evidence from different studies, and meta-research (cf.\ ICH E9 guideline \textit{Statistical Principles for Clinical Trials}; \citealp{ICH1998E9}). Naturally, some of these are more relevant for confirmatory research. Furthermore, comprehensive research protocols should be encouraged, particularly for research intended as confirmatory. We made a start in this direction with our list of essential items for study protocols for empirical methodological research in \hyperref[tab:checklistnew]{Table~\ref*{tab:checklistnew}}. Standardized reporting should be promoted in both exploratory and confirmatory research, possibly through the development of reporting guidelines.

And again in line with our recommendations for researchers in the previous subsection, journals should ideally require, or at least strongly encourage, authors to submit both the data, if possible, and the computational code used in the reported research together with their manuscript \citep{hofner2016biomj}. Editors should then have the computational reproducibility checked, as far as this is feasible, and the data and code made, ideally openly, available to readers. By now, the majority of leading statistics journals do encourage accessible data and code, but fall short of making it mandatory, and only very few perform reproducibility checks, such as the \emph{Biometrical Journal} and the \emph{Journal of the American Statistical Association}.

Beyond advancing the quality of reporting from their side, editors, funders, and other stakeholders should use their influence on what research will be done to achieve a more balanced and appropriate mix of studies for the current state of methodological research. Calls for proposals or special issues, as well as editorials, journal websites, and so on, could usefully distinguish between exploratory and confirmatory research, and, while valuing them equally, still focus on the type(s) of research that seem(s) most promising in their particular context. Looking at the state of the field, it would be very beneficial to recognize much more widely that articles proposing new methods are typically non-neutral and, importantly, to encourage research on existing methods, in particular neutral comparison studies \citep{norel2011self,Boulesteix2013neutral,Friedrich2024}. When confirmatory research projects are being promoted, as we are calling for, it could be worthwhile to incentivize their preregistration. Last but not least, meta-scientific projects examining methodological research itself should be funded. These can, for example, track changes due to the measures recommended here or diagnose the most prevalent current challenges to scientific progress in the field.


\section{Discussion}
\label{sec:discussion}

In the previous sections, we entertained that in most empirical methodological research exploration and confirmation are not clearly distinguished. In particular, we demonstrated with a survey that in many instances empirical methodological research can be considered what we call pseudo-confirmatory research. This places an epistemic burden on the fields and explains to some extent why parts of computational research are non-replicable and unreliable, and why we need more actual confirmatory methodological research to provide more reliable guidance to applied researchers who want to use these methods. In the previous section, we provided recommendations to improve the situation. 

However, while distinguishing these two modes of research is important conceptually, such a clear distinction between exploratory and confirmatory research is not always possible in practice, as already mentioned before. This will be illustrated in the following by two methodological studies coauthored by some of us. As a consequence, we think that it is practically more adequate to consider empirical research on a continuum between \lq\lq pure'' exploration and \lq\lq strict'' confirmation as \emph{idealized} end points 
\citep{Herrmann2024position, Hoefler2022, Jacobucci2022, Wagenmakers2012}. 

The first study investigated the behavior of the dimension reduction method UMAP and how it can be used to improve cluster detection using the density-based clustering method DBSCAN \citep{herrmannTopoClust2024}. It is an experimental study using both simulated and real data that is clearly exploratory in nature.\footnote{In general, confirmatory research in unsupervised learning appears to be specifically difficult because ``there is no [...] direct measure of success [and it] is difficult to
ascertain the validity of inferences drawn from the output of most unsupervised learning algorithms'' \citep[p.~487]{hastieElementsStatisticalLearning2009}.} Nevertheless, the conducted experiments were guided by certain but unspecific hypotheses about how the method works based on prior knowledge that has been established in previous studies, in particular the one that introduced the method UMAP \citep{mcinnes2020umap}. From these starting points, the experimental investigations contribute to a more nuanced understanding of the method UMAP and how it works given certain structure in data. It is demonstrated that it can be combined with the method DBSCAN to circumvent shortcomings including parameter sensitivity of the latter method and---maybe most importantly---the study provides further intuition and possible explanations as to why this is the case.

The second example is a benchmark study of survival prediction methods using multi-omics cancer data \citep{Herrmann2021}. The study was set up specifically to investigate the question of whether the different groups of omics data (e.g., copy number variation or gene expression) add predictive value compared to using only clinical data because prior research pointed into both directions. In addition, much of the study design (e.g., the datasets and variables to include, the methods to compare and their implementations) was specified prior to running the experiments since the goal of the study was to compare the methods as neutrally as possible. This research is certainly more confirmatory in nature than the first example, but it is still not confirmatory in a strict sense. In particular, the results of statistical tests conducted to evaluate differences in prediction performance were only reported with great caution and with an emphasis on their provisional and limited nature in this case. Moreover, while considerable effort had been put into specifying design and analysis options in advance based on prior research, it turned out that other design choices had to be made ad hoc. A follow-up study demonstrated that such design choices can have a considerable effect on the results of the benchmark experiment \citep{Niessl2022}. Nevertheless, the study provided important---partly unexpected---insights, in particular, it demonstrated that multi-omics data is not necessarily as useful as it has often been portrayed and that a simple Cox model using only clinical data is very competitive. Given the study design, the findings obviously cannot easily be generalized to other settings. This second example of methodological research may thus be considered \emph{rough} confirmatory data analysis (CDA), a type of investigation characterized by \citet{tukey1973exploratory} to lie between exploratory and confirmatory, and which \citet{Fife2022} recently specified as ``designed to evaluate specific hypotheses, though the hypotheses may not be ready for the rigor of strict CDA methods'' (p.~456). Note that an independent study corroborated some of the results in a (slightly) different experimental setup \citep{wissel2023systematic}.  

It should be noted that the concept of confirmatory research is tied to a set of important and interrelated epistemic questions, the discussion of which is beyond the scope of this paper. Some of these issues have been addressed in a related position paper on empirical research in ML coauthored by some of us \citep{Herrmann2024position} and other articles. For example, in a discussion for the special collection ``Towards Neutral Comparison Studies in Methodological Research'' in the \textit{Biometrical Journal}, \citet{Strobl2024} point out that asking \lq\lq \emph{Which method is best for a given set of datasets?}'' may be the wrong question to ask in the first place and \lq\lq argue that this research question implies assumptions which [they] do not consider warranted in methodological research'' (p.~1). Among other things, this is the case because it is very difficult to define the \emph{population of datasets} the results of a confirmatory study are intended to generalize to. This is a rather pronounced perspective on the important question of how far the results of a benchmark experiment in methodological research can be generalized beyond the specific experimental setup, which is the key goal of confirmatory research.

With that in mind, one may also wonder whether confirmatory methodological research could be theoretical, as the results of a theoretical analysis 
could be considered more general than empirical results based on specific datasets or data-generating mechanisms. We have deliberately focused on empirical methodological research because, in other disciplines, the concept of confirmatory research (and the exploratory--confirmatory distinction) has only ever been considered in reference to empirical research and is inherently linked to data. Nonetheless, we acknowledge that theoretical analyses can be valuable to examine how a method behaves under specific assumptions. On the one hand, such theoretical results have a broader scope in the sense that they apply to all datasets satisfying the stated assumptions, rather than only to the limited set of settings typically examined in an empirical study. However, this reliance on assumptions is also what can make such theoretical results less relevant in practical settings. The exact relationship between these two aspects is a very interesting question, but it goes beyond the scope of this paper.

Finally, especially in psychology and in the meta-scientific literature, some argue that a lack of background theory prevents specifying hypotheses a priori and, hence, that the field is not mature enough for strict confirmatory research\footnote{An illustrative example of a strict confirmatory study in this sense, that is, where the hypothesis being tested is derived from a strong (in this case, physical) theory and basically every experimental detail is prespecified (and preregistered), is the study by \citet{bartovs2025fair}, who show that \lq\lq [f]air coins tend to land on the same side they started'' (p.~1).} \citep{Fife2022}, which to some extent also holds for ML \citep{Herrmann2024position}.
While it is important to discuss these and related epistemic issues from a meta-scientific perspective, they should not distract from the fact that confirmatory research is possible---at least to an extent that makes it practically useful and valuable---even without a strong theory to derive hypotheses from. 
In medicine, too, there are sometimes no strong theories to support the hypotheses being tested. Nevertheless, phase III clinical trials are clearly prime examples of confirmatory research. The crucial point is that they are based on successive stages of prior (exploratory) investigations conducted in the different phases of medical research. In this sense, the field of medicine is certainly \lq\lq mature enough'' for confirmatory research even in cases without strong theories. In our opinion, this is not yet fully the case for empirical methodological research, and it should also mature in this sense. With this work, we hope to contribute to this improvement by raising awareness of the conceptual and practical differences between exploration and confirmation in empirical methodological research.


\section{Conclusion}
\label{sec:conclusion}

We suggest extending the concepts of exploratory and confirmatory research to empirical methodological research. 
Both types of research are vital to scientific advancement, particularly because progress in one supports the other, and recognizing the conceptual and practical differences between the two approaches will improve method development as well as guidance for applied researchers on which method to use in their specific use cases. 
In particular, we advocate that exploratory research should be clearly presented as such in order not to be perceived as confirmatory. Moreover, there should be progress towards more confirmatory methodological research and it should deserve more attention and recognition than it currently receives in the methodological literature and scientific community. 
We thus consider noteworthy the advice by \citet{drude2022planning}, formulated in the context of preclinical research, who \lq\lq recommend treating all research that does not explicitly state its confirmatory nature as exploratory'' (p.~8). To reiterate, this does not mean that it should be considered inferior research, but rather research that serves a different scientific goal than confirmatory research. 
In order to ensure the appropriate interpretation of results and to derive reliable guidance on how and when to use a specific \enlargethispage{\baselineskip}
method, it is crucial to communicate transparently which part of a study is exploratory or confirmatory and to what extent.

\section*{Funding Information}
The work was partly funded by the German Research Foundation (individual grants BO3139/7-2 and BO3139/9-1 to ALB).

\section*{Acknowledgments}
We thank Savanna Ratky for language editing and Christina Sauer, Maximilian Mandl, and Milena Wünsch for helpful comments.

\section*{Conflicts of Interest}
The authors declare no conflicts of interest.

\section*{Data Availability Statement}
The data that supports the findings of this study is available in the Supporting Information of this article.

\FloatBarrier

\printbibliography

\newpage
\end{document}